\documentclass[letterpaper, 10 pt, conference]{ieeeconf}  

\IEEEoverridecommandlockouts                              

\usepackage{mathptmx} 
\usepackage{times} 
\usepackage{amssymb}  
\usepackage{graphicx}
\usepackage{amsmath, nccmath}
\usepackage{tabularx,ragged2e,booktabs,caption}
\usepackage{geometry}
\usepackage{gensymb}

\title{\LARGE \bf
	Distributed Model Predictive Control based on Goal Coordination for Multi-zone Building Temperature Control 
}

\author{Roja Eini$^{1}$ and Sherif Abdelwahed$^{2}$
	\thanks{$^{1}$Roja Eini is PhD student of Department of Electrical Engineering,
		Virginia Commonwealth University, Richmond, VA, USA
		{\tt\small einir@vcu.edu}}%
	\thanks{$^{2}$Sherif Abdelwahed is faculty of Department of Electrical Engineering,
		Virginia Commonwealth University, Richmond, VA, USA
		{\tt\small sabdelwahed@vcu.edu}}%
}

\begin{document}

	\maketitle
	\thispagestyle{empty}
	\pagestyle{empty}

	\begin{abstract}
		In this paper, a distributed Model Predictive Control (DMPC) strategy is developed for a multi-zone building plant with disturbances. The control objective is to maintain each zone’s temperature at a specified level with the minimum cost of the underlying HVAC system. The distributed predictive framework is introduced with stability proofs and disturbances prediction, which have not been considered in previous related works. 
		The proposed distributed MPC performed with 48\% less computation time, 25.42\% less energy consumption, and less tracking error compared with the centralized MPC. The controlled system is implemented in a smart building test bed.
	\end{abstract}  
	
	\section{INTRODUCTION}
	
	Heating, ventilation and air conditioning (HVAC) systems in buildings currently account for 57\% of the US energy consumption, and 50\% of the world energy use. Thus, nowadays finding proper control systems in HVAC plants to reduce energy usage in the building sectors is of crucial importance.\\ \indent
	Model Predictive Control (MPC) strategy has been under significant attention for management of building plants recently [1-5]. However, using the centralized MPC approaches for controlling temperature of a multi-zone building system is impractical for several reasons. First, since there is a large number of system inputs and outputs for a building model with a large number of zones, it requires large centralized computational effort. Second, there are centralized modeling issues associated with global data collection and control actuation by a centralized agent. Third, when the centralized controller fails, the whole system is out of control and control integrity cannot be guaranteed.\\ \indent
	Distributed approaches come into play to deal with the above problems. The idea behind distributed approaches is to split the centralized system into subsystems whose control is assigned to a certain number of local controllers. Depending on the degree of interaction between the subsystems, the agents may need to communicate to coordinate themselves [6-7]. Several studies [8-10] have addressed the building temperature regulation problem using distributed MPC approach. In [8], a small temperature control model of a three-zone building does not include the open door situation and only one information per time step is being exchanged. In [9], a simplified two-masses system is considered as the plant model, but did not take into account the pressure and temperature dynamics. The authors in [10] focused on the modular distributed control of building temperature considering thermal and electrical energy sources, without considering the environment temperature predictions. None of these works considered the closed-loop system feasibility and stability.\\ \indent
	In this paper, a distributed MPC controller is designed and applied on a building's thermal model consisting of six rooms with the open/close door/window conditions. The objective is tracking the rooms' temperature set-points with minimum energy consumption considering the outdoor disturbances. The couplings between zones' states and inputs are also included in the model. Besides, a centralized MPC controller is developed and compared with the proposed distributed MPC. From the simulation results, the distributed MPC approach showed better performance in terms of desired set-point tracking, system cost minimization, and computation time compared with the centralized MPC. In contrast with previous works, this paper considers the disturbances predictions as part of the system model in the predictive approach and addresses the feasibility and stability aspect of the problem by the proposed coordination strategy. \\ \indent
	The rest of the paper is organized as follows. Section II introduces the models and parameters of the six-room system. Section III introduces the centralized and Distributed Model Predictive Control approaches. In the next section, the simulation results are shown. Section V provides a conclusion and discusses future research.
	
	\section{SYSTEM DEFINITION}
\begin{figure}[thpb]
		\begin{center}
			\framebox{
				\parbox{3in}
				{
					\centering
					\includegraphics[height=3cm, width=7cm]{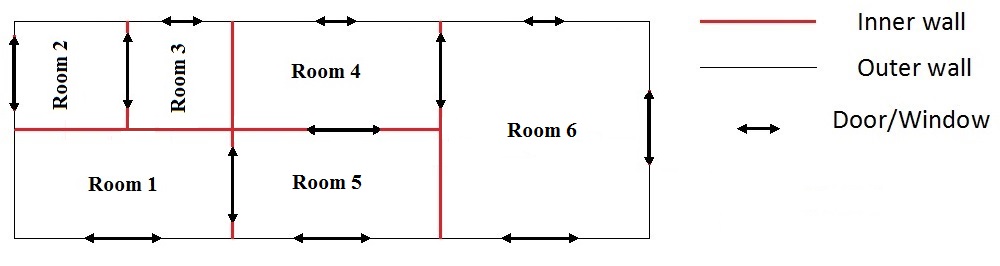}
				}
			}
			\caption{Six-room model plan}
			\label{fig:four_room}
		\end{center}
	\end{figure}

	The system considered is a building consisting of six rooms (subsystems) with the thermal exchange through the inner wall and inner door between them. The rooms also have thermal exchanges with the environment. The physical system layout is shown in Fig. \ref{fig:four_room} [11-12]. \\
	In Fig. \ref{fig:four_room}, the rooms are considered next to each other with all the windows and the exit door closed, and the internal doors between two rooms are open. Also, each room has one heater (AC). The control variables are the heater switches, temperature, and airflow settings. The general model can be developed based on the convection and conduction equations  (\ref{eq:air_flow}) and (\ref{eq:conduction}) for the heat transfer. \\
	\begin{equation}  \small
	\frac{dQ_{flow}}{dt}=M\times C_{air}(T_{adjacent}-T_{room})
	\label{eq:air_flow}
	\end{equation} 
	\begin{equation}    \small
	\frac{dQ_{conduction}}{dt}=\frac{(T_{adjacent}-T_{room})}{R}
	\label{eq:conduction}
	\end{equation}  \indent
	Table \ref{tab:system parameters description} parameters are used in this section's equations. Thus, the thermal exchange rate of a convection-based component $q_{f,i}$ is described based on (\ref{eq:air_flow}) and the thermal exchange rate of the conduction-based component $q_{c,j}$ is described based on (\ref{eq:conduction}). Then, the room heat exchange would be described as (\ref{eq:Q_room_convection_conduction}). \\
	\begin{table}
		\begin{center}
			\caption{System Parameters Description}
			\begin{tabular}{| c | m{6cm}|}
				\hline
				M & {amount of air flowing between two regions (in Kg/h)} \\  \hline
				$C_{air}$ & {heat capacity of air (J/Kg \degree C)} \\ \hline
				$T_{adjacent}$ & {the temperature of the region exchanging airflow with adjacent component (\degree C)} \\ \hline
				$T_{room}$ & {the ambient temperature of the room (\degree C)} \\ \hline
				$R$ & {thermal resistance of a given component} \\ \hline
				$Q_{flow}$ & {heat flow due to convection (J)} \\ \hline
				$Q_{conduction}$ & {heat flow due to conduction (J)} \\ \hline
				$Q_{room}$ & {total heat amount in the room (J)} \\ \hline
				$F$ & {set of all the convection-based components} \\ \hline
				$C$ & {set of all the conduction-based components} \\ \hline
			\end{tabular}
			\label{tab:system parameters description}
		\end{center}
	\end{table}
	\begin{table}
		\begin{center}
			\caption{System Parameters Numerical Values}
			\begin{tabular}{|c|c|}
				\hline
				C & 1005.4              \\  \hline
				$m_i, \,\,\,\, i=1,\cdots,6$ & 102.0425 \\ \hline
				$T_{room2-3}, T_{room1-5}, T_{room4-5}, T_{room4-6} (initial)$  & 10        \\ \hline
				$M_{indoori}$ & 20 \\ \hline
				$R_{indoors}$ & 0.000208 \\ \hline
				$R_{walls-ini}$ & 0.0000696 \\ \hline
				$R_{walls-outi}$ & 0.0000321  \\ \hline
				$M_{outdoori}, M_{window3}$ , $M_{window4}$ , $M_{window6}$& 35 \\ \hline
				$R_{outdoors}$ & 0.000208  \\ \hline
				$R_{window3}$ , $R_{window4}$ , $R_{window6}$ & 0.0000593542 \\ \hline
				$wc_{window3}$ , $wc_{window4}$ , $wc_{window6}$ & 1 \\ \hline
				$wf_{window3}$ , $wf_{window4}$ , $wf_{window6}$ & 0 \\ \hline
				$wc_{outdoor1}$ , $wc_{outdoor2}$ , $wc_{outdoor5}$, $wc_{outdoor6}$ & 1 \\ \hline
				$wf_{outdoor1}$ , $wf_{outdoor2}$ , $wf_{outdoor5}$,  $wf_{outdoor6}$ & 0 \\ \hline
				$wc_{indoori}$ & 0 \\ \hline
				$wf_{indoori}, wf_{aci}$ & 1 \\ \hline
			\end{tabular}
			\label{tab:system parameters numerical values}
		\end{center}
	\end{table}
	
	\begin{equation}   \small
	\frac{dQ_{room}}{dt}=\sum\limits_{i\in F}^{}{w_{f,i} q_{f,i}}+\sum\limits_{j\in C}^{}{w_{c,j} q_{c,j}}
	\label{eq:Q_room_convection_conduction}
	\end{equation}   \\ \indent
	In (\ref{eq:Q_room_convection_conduction}), $w_{f,i}$ is the binary weight and can be $0$ or $1$; $0$ when the door/window is closed and $1$ otherwise.  $w_{c,j}$ is also a binary weight and can be $0$ or $1$; $1$ when the door/window is closed and $0$ otherwise.\\ \indent
	Eventually, (\ref{eq:general_heat excahnge_two room}) illustrates the total heat exchange in a general format for each room. \\
	\begin{equation}  \small
	\frac{dT_{room}}{dt}=\frac{1}{m_{i}\times C_{air}}\times \frac{dQ_{room}}{dt}
	\label{eq:general_heat excahnge_two room}
	\end{equation} 
	\indent
	Therefore, the temperature of room $i, \,\,\,\,\,\, i=1, \cdots, 6$ in general is built as (\ref{eq:roomi_four room}). \\
	
	\begin{fleqn}
		\begin{equation}  \small
		\begin{aligned}[b]
		&\frac{d{x_{i}}}{dt} =(\frac{1}{m_i\times C})[ \frac{T_o-x_i}{R_{walls-outi}} + \frac{T_{roomi-j}-x_i}{R_{walls-ini}} +wc_{outdoori} \frac{T_o-x_i}{R_{outdoori}} \\
		&+ wf_{outdoori} M_{outdoori} C (T_o-x_i)+ wc_{indoori} \frac{T_{roomi-j}-x_i}{R_{indoori}} \\
		&+wf_{indoori} M_{indoori} C (T_{roomi-j}-x_i) +{wc_{windowi}}{\frac{(T_o-x_i)}{R_{windowi}}}\\
		&+ wf_{windowi}  M_{windowi}  C  (T_o-x_{i}) +wf_{ac} M_{ac} C (T_{ac}-x_i)]
		\end{aligned}
		\label{eq:roomi_four room}
		\end{equation}
	\end{fleqn}
	\normalsize  \indent
	In (\ref{eq:roomi_four room}),  $x_i$ is the temperature of room $i$, and $T_o$ is the outside temperature as the disturbance. $R_{indoors}$, $R_{outdoors}$, $R_{walls-outi}$, and $R_{walls-ini}$ are the total heat resistivity of indoors, outdoors, and the heat resistivity of walls from the outside and inside layers of room $i$, respectively. $wc_{indoor}$ and $wc_{outdoor}$ are the conduction weight between two rooms, and the conduction weight between the rooms and outside, respectively. Also, $T_{roomi-j}$ is the heat exchange between room $i$ and $j$. $M_{outdoor}$ , $M_{indoor}$ , and $M_{window}$ are the amount of airflow from outside to inside, the amount of airflow indoors, and the amount of airflow from the windows respectively. $M_{ac}$ is the amount of airflow of the heater. The numerical values used for the system simulations are stated in Table \ref{tab:system parameters numerical values}; $i=1,2, \cdots, 6$. Note that the model is developed modular to be modifiable if the system components' status change.
	\section{THE MODEL PREDICTIVE APPROACH}
	A centralized Model Predictive Controller is first applied on the model to compare its results with the proposed distributed MPC approach.
	\subsection{Centralized MPC}
	The state space model MPC approach is chosen as the centralized MPC framework [5]. By discretization of the system state space model (\ref{eq:roomi_four room}), the states predictions $P$ steps ahead of the current time are stated as (\ref{central_statepredict}).\\
	\begin{fleqn}
		\begin{equation} \small
		\begin{aligned}[b] 
		{\hat{x}(k|k)}&={A}{{\hat{x}}(k|k-1)}+Bu(k-1)+Ed(k-1)+L({\hat{y}}(k)-{\hat{y}}(k|k-1))\\
		{Y(k|k)}&=TH{\hat{x}(k|k)}+TG{\Delta{U}(k|k)}+TFu(k-1)+TVW(k|k)
		\end{aligned}
		\label{central_statepredict}
		\end{equation}
	\end{fleqn}   \indent
	In (\ref{central_statepredict}), $A$, $B$, and $E$ are the state space representation matrices of the system. $d$ is the disturbance, and $Y$, $G$, $T$, $F$, $H$, $V$, $W$ are defined as follows. \\
	\begin{equation}  \nonumber  \small
	{Y^T}(k)=
	\begin{bmatrix} 
	y(k+1|k)^T & y(k+2|k)^T  & \cdots & y(k+P|k)^T 
	\label{central_cost_Y}
	\end{bmatrix}
	\end{equation}

	\begin{equation} \nonumber  \small
	\begingroup 
	\setlength\arraycolsep{0.5pt}   
	G=
	\begin{bmatrix} 
	B & 0 & \cdots & 0\\
	(A+I)B &B & \cdots & 0\\
	\vdots & \vdots &\cdots & \vdots\\
	\sum\limits_{i=1}^{M}{A^{i-1}}B & \sum\limits_{i=1}^{M-1}{A^{i-1}}B & \cdots & B\\
	\sum\limits_{i=1}^{M+1}{A^{i-1}}B & \sum\limits_{i=1}^{M}{A^{i-1}}B & \cdots & (A+I)B\\
	\vdots & \vdots &\cdots & \vdots\\
	\sum\limits_{i=1}^{P}{A^{i-1}}B & \sum\limits_{i=1}^{P-1}{A^{i-1}}B & \cdots & \sum\limits_{i=1}^{P-M+1}{A^{i-1}}B
	\label{central_cost_G}
	\end{bmatrix}
	\endgroup
	T= 
	\begin{bmatrix} 
	C & 0 & \cdots & 0\\
	0 & C & \ddots & \vdots\\
	\vdots & \ddots &\ddots & 0\\
	0 & \cdots & 0 & C
	\end{bmatrix}
	\end{equation}\\
	\begin{equation}  \nonumber   \small
	{F^T}=
	\begingroup 
	\setlength\arraycolsep{1pt} 
	\begin{bmatrix} 
	B & (A+I)B & \cdots & \sum\limits_{i=1}^{M}{A^{i-1}}B & \sum\limits_{i=1}^{M+1}{A^{i-1}}B & \cdots & \sum\limits_{i=1}^{P}{A^{i-1}}B
	\label{central_cost_F}
	\end{bmatrix}
	\endgroup
	\end{equation}    
	
	\begin{equation}  \nonumber  \small
	\begingroup 
	\setlength\arraycolsep{0.5pt}     
	{H^T}=
	\begin{bmatrix} 
	A & {A^2} & \cdots & {A^P}
	\label{central_cost_H}
	\end{bmatrix}
	\endgroup  \,\,\,\,\,\,\,\,\,\,\,\,\,
	V=
	\begin{bmatrix} 
	E & 0 & \cdots & 0\\
	AE & E & \cdots & 0\\
	\vdots & \vdots &\cdots & \vdots\\
	{A^{p-1}}E & {A^{p-2}}E & \cdots & E
	
	\end{bmatrix}
	\end{equation}
	
	\begin{equation}  \small
	{W^T}(k)=
	\begin{bmatrix} 
	{{d(k|k)}^T} & {{d(k+1|k)}^T} & \cdots & {{d(k+P-1|k)}^T}
	\label{central_cost_W}
	\end{bmatrix}
	\end{equation}  \\
	\indent  \normalsize
	Moreover, the cost function $J(k)$ in (\ref{central_cost}) penalizes deviations of the predicted outputs $\hat{y}(k+i|k)$ from a reference trajectory ${y_r}(k+i|k),\,\,\,\,\,\,\,i=1,2,\cdots , P$.
	\begin{fleqn}  
		\begin{equation} \small
		\begin{aligned}[b] 
		{J}(k)&=\sum\limits_{i=1}^{P}{{\lVert{({{\hat{y}(k+i|k)}-{y_r}(k+i|k)})}\lVert}_{Q}^2}+\sum\limits_{i=1}^{M}{{\lVert {\Delta {u}(k+i-1|k)}\lVert}_{R}^2} 
		\end{aligned}
		\label{central_cost}
		\end{equation}
	\end{fleqn} 
	\normalsize   \indent
	In (\ref{central_cost}), $P$ and $M$ are the prediction and the control horizons respectively. ${Q}$ and ${R}$ are the weight matrices, ${y_r}$ and ${\Delta {u_i}}$ are the reference trajectory and the input increment vector respectively. Thus, to minimize the cost function (\ref{central_cost}) subject to the system model description and the prediction equations, the centralized MPC algorithm would be as follows. 
	\\
	\indent 
	Step 0: Get the system model at the current time. \\ \indent \normalsize
	Step 1: at k=0; measure $y(0)$, determine ${Y_r}(0)$ and solve the optimization problem to calculate $\Delta{u}(0)$, and then substitute ${u}(0)$ by the calculated $\Delta{u}(0)$.\\ \indent  \normalsize
	Step 2: at time k$>$0; measure $y(k)$, determine ${Y_r}(k)$ and solve the optimization problem to calculate $\Delta{u}(k)$, and then substitute ${u}(k)$ by the calculated $\Delta{u}(k)$.\\ \indent \normalsize
	Step 3: $k=k+1$ and go back to step 2.\\ \indent  \normalsize
	Hence, the whole system is monolithic using the centralized MPC, and only one MPC controller is assigned to the system. Therefore, there is one complicated large optimization problem with various variables being calculated at each time step.  \normalsize
	
	\subsection{Distributed MPC}  
\begin{figure}[thpb]
		\begin{center}
			\framebox{\parbox{3in}{
					\centering
					\includegraphics[height=4cm, width=6cm]{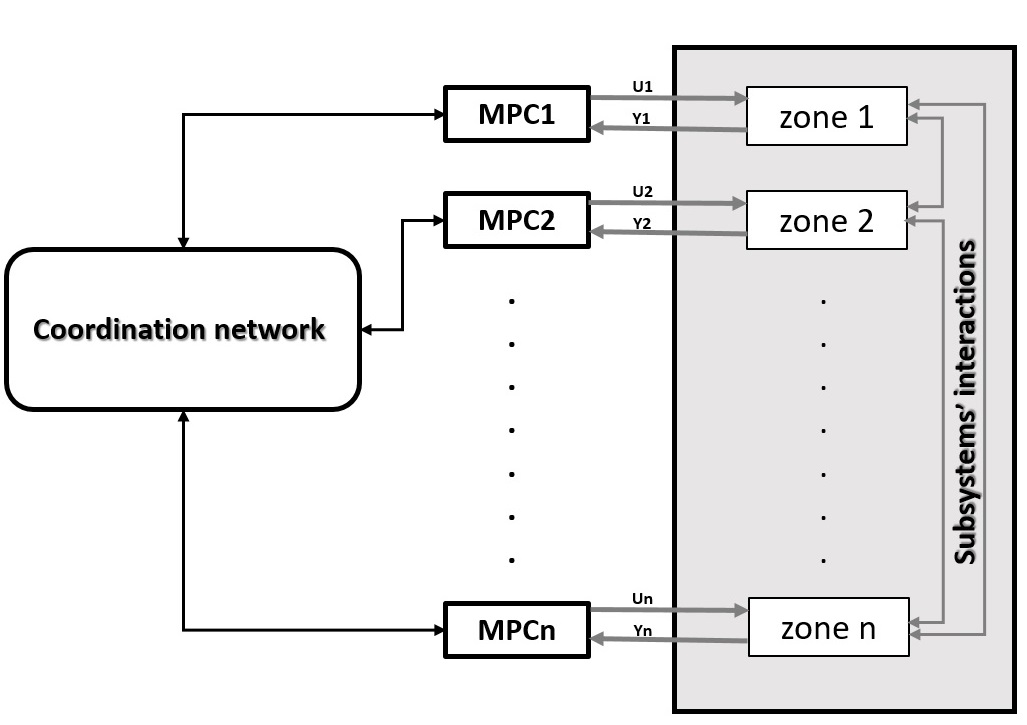}
			}}
			\caption{Distributed MPC of n interacted subsystems with exchanging information}
			\label{fig:blockdiagram_roja}
		\end{center}
	\end{figure}

\normalsize
	Distributed MPC approach is known to be effective specifically in building management systems since large building plants consist of various subsystems. Using a distributed MPC, one or more local MPCs are assigned to each subsystem of the whole plant, and they coordinate together to achieve a specific global performance of the entire system [13-15]. A simple block diagram of distributed MPC for a plant with $n$ zones (subsystems) is presented in Fig. \ref{fig:blockdiagram_roja}.
	\normalsize
	  \normalsize
	The distributed MPC algorithm proposed in this paper considers not only the future output and manipulated input predictions of the neighbor zones but also the disturbances predictions in each local controller. The goal is to attain a satisfactory global performance of the entire system with minimum computation demand [8].\\ \indent 
	To begin with, MPC controllers and the coordination mechanism are required to be specified. Each MPC controller itself has composed of three prominent parts: a state predictor, an interaction predictor, and an optimizer. Each local MPC controller has one or more objective function as (\ref{general_cost_func}) containing the tracking error (between the future output ($\hat{y_i}$) and the determined reference signal ${{y_i}^d}$) and the control effort increments ($\Delta{u}$). 
	
	\begin{fleqn}  
		\begin{equation}  \small
		\begin{aligned}[b] 
		{J_i}(k)&=\sum\limits_{l=1}^{P}{{\lVert{({{{\hat{y}}_i}(k+l)-{{y_i}^d}(k+l)})}\lVert}_{Q_i}^2}+\sum\limits_{l=1}^{M}{{\lVert {\Delta {u_i}(k+l-1)} \lVert}_{R_i}^2}\\
		\end{aligned}
		\label{general_cost_func}
		\end{equation}
	\end{fleqn} \indent
	\normalsize
	In (\ref{general_cost_func}), $P$ and $M$ are the prediction and the control horizons respectively. ${Q_i}$ and ${R_i}$ are the weight matrices, ${{y_i}^d}$ and ${\Delta {u_i}}$ are the reference trajectory and the input increment vector of subsystem $i$ respectively. ${{y_i}^d}$ is obtained by a smooth approximation from the current value of output ${y_i}(k)$ towards the known reference ${r_i}(k)$ by (\ref{w_equ}).  \\
	
	\begin{equation} \label{w_equ}  \small
	\begin{split}
	&{{y_i}^d}(k)={y_i}(k),\,\,\,\,\,\,\,{{y_i}^d}(k+l)={\alpha_i}{{w_i}(k+l-1)}+(1-{\alpha_i}){r_i}(k+l)\\
	&l=1 \cdots P
	\end{split}
	\end{equation}
	\indent
	\normalsize
	To obtain the values of future control laws ${u_i}(k+l|k)$ for each subsystem $i$, the local objective function of (\ref{general_cost_func}) should be minimized at each time step $k$. Then, the global objective function of the whole plant at each time step $k$ would be defined as (\ref{general_general_cost_func}). 
	
	\begin{fleqn}
		\begin{equation} 
		{J}(k)=\sum\limits_{i=1}^{m}{J_i}(k) \\
		\label{general_general_cost_func}
		\end{equation}
	\end{fleqn}  \\ \indent
	\normalsize
	In the above, $m$ is the total number of subsystems. The values of the system's predicted outputs and states are calculated through (\ref{state_prediction}), and then substituted in the cost function. 
	
	\begin{fleqn}  \small
		\begin{equation}
		\begin{aligned}[b] 
		{\hat{x}_i}(k+l|k)&={{A_{ii}}^l}{{\hat{x}_i}(k|k)}+\sum\limits_{s=1}^{l}{{A_{ii}}^{s-1}{B_{ii}}{u_i}{(k+l-s|k)}}\\
		&+\sum\limits_{s=1}^{l}{{A_{ii}}^{s-1}{\hat{w_i}}{(k+l-s|k-1)}} \\
		{\hat{y}_i}(k+l|k)&={{C_{ii}}}{{\hat{x}_i}(k+l|k)}+{{\hat{v}_i}(k+l|k-1)}
		\end{aligned}
		\label{state_prediction}
		\end{equation}
	\end{fleqn} \\
	\indent
	\normalsize
	Moreover, the states and inputs interaction equations between the subsystems can be stated as (\ref{interactions}). Note that the disturbances are included in the input vector of the model.
	
	\begin{fleqn}  
		\begin{equation} \small
		\begin{aligned}[b] 
		{w_i}(k)&=\sum\limits_{j=1}^{m}{{A_{ij}}{{x_j}(k)}}+\sum\limits_{j=1}^{m}{{B_{ij}}{{u_j}(k)}}
		\,\,\,\,\,\,\,\,\,
		{v_i}(k)=\sum\limits_{j=1}^{m}{{C_{ij}}{{x_j}(k)}}\\
		\end{aligned}
		\label{interactions}
		\end{equation}
	\end{fleqn}

	\indent
	\normalsize
	Dual decomposition method based on Lagrangian function is proposed through the coordination mechanism in this work. The idea is to impose the interconnecting constraints into the objective function by the Lagrangian multipliers and approximately solve the dual cost function (duality theory is explained in [5]).
	Thus, the augmented function for each subsystem $i$ is stated as (\ref{augmented_cost}).
	
	\begin{fleqn}
		\begin{equation}  \small
		\begin{aligned}[b] 
		{L}_{i}(k)&= {J_i}(k) + \sum\limits_{j\in N_i}{} [{\lambda_{ij}(k)}^T (z_{ij}(k)-\hat{z}_{ij}(k)) + \frac{\rho}{2}  {\lVert z_{ij}(k)-\hat{z}_{ij}(k) \lVert_{2}^2}]  \\
		z_{ij}(k) &= [{w_{ij}(k)} \,\,\,\,\, {v_{ij}}(k)]^T  \,\,\,\,\,\,\,\,\,\,\,\,\,\,\,\,
		N_i : Neighborhood \,\, of \,\, subsystem \,\, i
		\end{aligned}
		\label{augmented_cost}
		\end{equation}
	\end{fleqn}
	\indent
	\normalsize
	Note that the subsystems only communicate with their neighbors in $N_i$ to get the interconnecting information from them and to send their last updated variables (states and inputs) to them at each time step.
	Therefore, the optimization problem for each local controller is the minimization of (\ref{augmented_cost}) subject to constraints (\ref{state_prediction}). Consider that the duality function coefficients $\lambda$ and $\rho$ should be optimized as well as the input signal in each iteration. $\lambda$ is updated through (\ref{lambda}). \\
	\begin{fleqn}
		\begin{equation} \small
		\begin{aligned}[b] 
		\lambda_{ij}^{s+1}(k) &= \lambda_{ij}^{s}(k)+ \alpha ^{s} (z_{ij}^s(k)-{\hat{z}_{ij}^s(k)})
		\end{aligned}
		\label{lambda}
		\end{equation}
	\end{fleqn} \\   \indent 
	Defining the following matrices; \\
	\small
	\begin{equation}  \nonumber
	\begingroup 
	\setlength\arraycolsep{0.5pt}  
	{{\tilde A}_i}= 
	\begin{bmatrix} 
	{diag_P}\{{A_{i,1}}\} & \cdots & {diag_P}\{{A_{i,i-1}}\} & 0 & {diag_P}\{{A_{i,i+1}}\} & \cdots & {diag_P}\{{A_{i,m}}\}
	\label{solutions1}
	\end{bmatrix}
	\endgroup
	\end{equation}
	\small
	\begin{equation}  \nonumber
	\begingroup 
	\setlength\arraycolsep{0.5pt}  
	{{\tilde B}_i}= 
	\begin{bmatrix} 
	{diag_P}\{{B_{i,1}}\} & \cdots & {diag_P}\{{B_{i,i-1}}\} & 0 & {diag_P}\{{B_{i,i+1}}\} & \cdots & {diag_P}\{{B_{i,m}}\}
	\label{solutions2}
	\end{bmatrix}
	\endgroup
	\end{equation}
	\small
	\begin{equation}  \nonumber  \small
	\begingroup 
	\setlength\arraycolsep{0.5pt}  
	{{\tilde C}_i}= 
	\begin{bmatrix} 
	{diag_P}\{{C_{i,1}}\} & \cdots & {diag_P}\{{C_{i,i-1}}\} & 0 & {diag_P}\{{C_{i,i+1}}\} & \cdots & {diag_P}\{{C_{i,m}}\}
	\label{solutions3}
	\end{bmatrix}
	\endgroup
	\end{equation}
	\small
	\begin{equation}  \small
	{{\tilde \Gamma}_i}=
	\begingroup 
	\setlength\arraycolsep{0.5pt}   
	\begin{bmatrix} 
	{0}_{(M-1){{n_{ui}}*{n_{ui}}}} & I_{(M-1)*{n_{ui}}}\\
	{0}_{{n_{ui}}*(M-1){n_{ui}}} & I_{n_{ui}}\\
	\vdots & \vdots\\
	{0}_{{n_{ui}}*(M-1){n_{ui}}} & I_{n_{ui}}
	\label{solutions4}
	\end{bmatrix}
	\endgroup
	\,\,\,\,\,\,\,\,\, {\tilde \Gamma}=   
	{diag} \{ {{\tilde \Gamma}_1} \cdots {{\tilde \Gamma}_m}\}
	\,\,\,\,\,\,\
	{\tilde {B_i}}=   
	\tilde{{\tilde {B_i}}} {\tilde \Gamma}
	\end{equation} 
	
	\normalsize  
	\noindent
	Interaction predictions of subsystem $i$  would be as (\ref{WiandViprediction}). \\
	
	\begin{fleqn}
		\begin{equation}  \small
		\begin{aligned}[b]   
		{\hat {W_i}(k,P|k-1)}&=\tilde{A_i}{\hat{X}}(k,P|k-1)+\tilde{B_i}{U}(k-1,M|k-1)\\
		{\hat {V_i}(k,P|k-1)}&=\hat{C_i}{\hat{X}}(k,P|k-1)  
		\end{aligned}
		\label{WiandViprediction}
		\end{equation}
	\end{fleqn}   \\ \indent
	The state and output predictions would be defined as (\ref{stateandoutputpredictions}).
	
	\begin{fleqn}
		\begin{equation} 
		\begin{aligned}[b]   
		{\hat {X_i}(k+1,P|k)}&=\bar{S_i}[\bar{A_i}\hat{x_i(k|k)}+\bar{B_i}U_i(k,M|k)+\hat{W_i}(k,P|k-1)]\\
		{\hat {Y_i}(k,P|k-1)}&=\bar{C_i}[\hat{X_i}(k+1,P|k)+T_i\hat{V_i}(k,P|k-1)]  
		\end{aligned}
		\label{stateandoutputpredictions}
		\end{equation}
	\end{fleqn}  
	\noindent
	where matrices $T_i$, $\bar{S_i}$, $\bar{A_i}$, $\bar{B_i}$, and $\bar{C_i}$ are stated as (\ref{Cibar}).
	
	\small
	\begin{equation}  \nonumber
	{T_i}=  
	\begin{bmatrix} 
	0_{(P-1){n_{yi}}*{n_{yi}}} & I_{(P-1){n_{yi}}}\\
	0_{{n_{yi}}*(P-1){n_{yi}}} & I_{n_yi}
	\label{Ti}
	\end{bmatrix}  \,\,\,\,\,\,\,\,\
	{\bar{S_i}}= 
	\begin{bmatrix} 
	{A_{ii}}^0 & \cdots & 0 \\
	\vdots & \ddots & \vdots \\
	{A_{ii}}^{P-1} & \cdots & {A_{ii}}^0
	\end{bmatrix}
	\end{equation} 
	\small
	\begin{equation}  \nonumber  \small
	{\bar{A_i}}= 
	\begin{bmatrix} 
	{A_{ii}} \\
	0_{Pn_{yi}*n_{yi}}
	\label{Aibar}
	\end{bmatrix}   \,\,\,\,\,\,\,\,\
	{\bar{B_i}T}= 
	\begin{bmatrix} 
	{diag_M\{B_{ii}}\} & & &\\
	0_{n_{ui}} &\cdots & 0_{n_{ui}} & B_{ii} \\
	\vdots & \ddots & \vdots & \vdots \\
	0_{n_{ui}} & \cdots & 0_{n_{ui}} & B_{ii}
	\end{bmatrix}
	\end{equation} 
	\begin{equation}  \small
	{\bar{C_i}}= diag_P\{ C_{ii} \}
	\label{Cibar}
	\end{equation} \\
	\normalsize  \indent
	Therefore, the control solution for the optimization problem would be as (\ref{controllaw}). \\
	\begin{equation} \small
	{U_i(k,M|k)} = {{\Gamma_1}^\prime}u_i(k-1)+\bar{\Gamma_i}\bar{K_i}[Y_i^d(k+1,P|k)-\hat{Z_i}(k+1,P|k)]
	\label{controllaw}
	\end{equation}
	\noindent
	where $K_i$, and $\hat{Z_i}$ are defined in (\ref{controllaw_coeffs}). \\
	
	\begin{fleqn}
		\begin{equation}  \nonumber   \small
		\begin{aligned}[b]  
		{\hat{Z_i}(k+1,P|k)}&= S_i[\bar{B_i}{\Gamma_i}^\prime u_i(k-1)+\bar{A_i}\hat{x_i}(k|k)+\hat{W_i}(k,P|k-1)] \\
		&+T_i\hat{V_i}(k,P|k-1)
		\end{aligned}
		\label{zihat}
		\end{equation}
	\end{fleqn}  \\
	\small
	\begin{equation}  \nonumber  
	{\bar{K_i}}= {H_i}^{-1} {{N_i}^T} {\bar{Q_i}}  
	\,\,\,\,\,\,\,\,\,\,\,\,\,\,\,
	H_i={{N_i}^T}{\bar{Q_i}}{N_i}+{\bar{R_i}}
	\,\,\,\,\,\,\,\,\,\,\,\,\,\,\,
	S_i={\bar{C_i}}{\bar{S_i}}
	\label{Si}
	\end{equation}
	\small
	\begin{equation}  \nonumber  \small
	\bar{Q_i}=diag_P\{Q_i\}
	\,\,\,\,\,\,\,\,\,\,\,\,\,\,\,
	\bar{R_i}=diag_P\{R_i\}
	\,\,\,\,\,\,\,\,\,\,\,\,\,\,\,
	N_i={S_i}{\bar{B_i}}{\bar{\Gamma _i}}
	\label{Ni}
	\end{equation} 
	\small
	\begin{equation} \label{controllaw_coeffs} \small
	{{\Gamma_i}^\prime}= 
	\begin{bmatrix} 
	I_{n_{ui}}\\
	\vdots\\
	I_{n_{ui}}
	\end{bmatrix}   \,\,\,\,\,\,\,\,\,\,\,\,\,\
	{\bar{\Gamma_i}}= 
	\begin{bmatrix} 
	I_{n_{ui}} & \cdots & 0\\
	\vdots & \ddots & \vdots \\
	I_{n_{ui}} & \cdots & I_{n_{ui}} 
	\end{bmatrix}
	\end{equation}  
	\normalsize \\  \indent
	Thus, the following distributed MPC strategy is proposed for the building plant.
	
	Step 1: 
	\begin{itemize} \nonumber
		\item Send $U_i(k-1,M|k-1)$ and $\hat{X}_i(k,P|k-1)$ to its neighboring controller $C_j$ (coordination mechanism). 
		\item Estimate the future state trajectories $\hat{X}_j(k,P|k-1)$ and control inputs $U_j(k-1,M|k-1)$ from its neighboring controller through information exchange (Goal coordination method [5]).
		\item Determine the desired trajectory $Y_d(k+1,P|k)$ based on MPC's configuration.
		\item Observe the values of $x_i(k)$.
		\item Build the $\hat{x}(k,P|k-1)$ and $U(k,P|k)$ by adding the subsystem's state estimations $\hat{x}_i(k,P|k-1)$ and control input $U_i(k,P|k)$, and the subsystem's neighbor information $\hat{x}_j(k,P|k-1)$ and $U_j(k-1,M|k-1)$ to attain the predictions of $\hat{W_i}(k,P|k-1)$ and $\hat{V_i}(k,P|k-1)$ (from (\ref{WiandViprediction})).
	\end{itemize}
	
	Step 2:  
	\begin{itemize} 
		\item Calculate the optimal control law $U_i(k,M|k)$ from (\ref{controllaw}).
		\item Apply the first element of the optimal control $U_i(k,M|k)$ ($u_i(k)$) to the system.
		\item Update $\lambda$ from (\ref{lambda}).
	\end{itemize}
	
	Step 3: 
	\begin{itemize} 
		\item Compute the estimation of the future state trajectory of $i$th subsystem over the horizon $P$ from (\ref{stateandoutputpredictions}).
	\end{itemize}
	
	Step 4:  
	\begin{itemize} 
		\item change $k$ to $k+1$ and go back to step 1 and repeat the algorithm.
	\end{itemize} 
	\indent
	
	The coordination strategy in the proposed distributed algorithm based on goal coordination avoids global communication in the whole network and enhances the closed-loop system stability and feasibility [5]. Assuming the existence of a feasible input sequence for each subsystem $i$ at k=0, the optimization problem has a feasible solution for each subsystem $i$ at all $k\geq 0$. For stability analysis, (\ref{Lyapunov}) is defined as the Lyapunov function which will be solved off-line. \\
	\begin{equation}  \nonumber  \small
	A^TPA-P=-F, \,\,\,\,\,\,\,\,\,\,\,\, P=
	\begin{bmatrix} 
	P_{11} & P_{12} & \cdots & P_{1m} \\
	P_{21} & P_{22} & \cdots & P_{2m} \\
	\vdots & \vdots & \ddots & \vdots \\
	P_{m1} & P_{m2} & \cdots & P_{mm}
	\end{bmatrix}   
	\end{equation} 
	\begin{equation}   \small
	F= diag(F_1,F_2,\cdots,F_m),  \,\,\,\,\,\,\, F_i(0)=F_i(1)=\cdots=F_i(N-1)=F_i
	\label{Lyapunov}
	\end{equation}  \indent
	Having relationship (\ref{stability_proof}) from [5],
	\begin{equation} \small
	J(k)(x(k))\leq J(0)(x(0))- \sum\limits_{k=0}^{K-1} \sum\limits_{i=0}^{m} J_i(k)(x_i(k),u_i(k))\leq J(0)(x(0))
	\label{stability_proof}
	\end{equation} 
	(\ref{stability_proof_2}) is attained.
	\begin{fleqn}
		\begin{equation}    \small
		\begin{aligned}[b]
		\frac{1}{2}\lambda _{min}(F){\lVert x(k)\lVert}^2  &\leq J(k)(x(k)) \\
		J(k)(x(k))\leq J(k)(x(0))=\frac{1}{2}{x(0)}^TPx(0) &\leq \frac{1}{2} \lambda_{max}(P){\lVert x(0)\lVert}^2 
		\label{stability_proof_2}
		\end{aligned}
		\end{equation}
	\end{fleqn}  \\
	
	Thus, it is proved that $\lVert x(k)\lVert \leq \sqrt{\frac{\lambda_{max}(P)}{\lambda _{min}(F)}} {\lVert x(0)\lVert} $, which shows that the closed-loop system is asymptotically stable under the distributed algorithm.
	\\ \indent
	Moreover, the disturbances predictions are considered as part of the system model through the proposed algorithm. In result, less computation and complication, and better overall performance are attained using the proposed framework.  
	\section{SIMULATION AND RESULTS}
	The desired trajectory for each room's temperature is between 5 to 25 \degree C in 5 different time periods (0-6 AM, 6-12 AM, 12-6 PM, 6-9 PM, and 9-12 PM) regarding the occupancy condition. To maintain occupant comfort, temperature set points during occupied hours (0-6 AM and 6-12 PM) are higher than the vacant periods (6-12 AM and 0-6 PM). The environment temperatures as the disturbances are also assumed to be between -6 to 4 \degree C.\\ \indent
	In the centralized case, the discrete-time state space equation of the system is considered and the cost function is a monolithic global function as (\ref{globalcost}) for the whole system.
	\begin{equation} \small
	J(k)=\sum\limits_{i=1}^{6}\{ {{\lVert{({x_i(k)}-{{x_i}^d(k)})}\lVert}_{Q_i}^2}+{{\lVert{{u_{i}(k)}}\lVert}_{R_i}^2} \}
	\label{globalcost}
	\end{equation} \\
	where $x_i$s and ${x_i}^d$s are the rooms' temperatures and the reference trajectories of room $i$ respectively. Note that the control variables for centralized MPC are $T_{ac}i$, $M_{ac}i$, and $wf_{ac}i$.\\ \indent
	The diagonal weighting matrices $Q_i$ and $R_i$ are chosen as (\ref{QandR}), associated with the set-point errors' and inputs' weightings respectively.
	\begin{equation} \small
	Q_i=1.5\times {I_{12\times12}}, \,\,\,\,\,\,\,\,R_i=(\frac{1}{1600})\times {I_{6\times6}}
	\label{QandR}
	\end{equation} \\ \indent
	At each time instant, the optimization problem is solved using a MATLAB optimization solver, then the optimum input is considered as the current input for the next step. The attained input at each step also builds the actual system output $y$. \\ \indent
	In the distributed case, six subsystems with their own objective functions and at least one neighbor share their inputs, disturbances, and states information with their neighbors. \\ \indent
	Thus, the equation for subsystem $i, \,\,\,\, i=1,2,\cdots, 6$  is (\ref{eq:roomi_four room}), with objective function (\ref{objectivei}). 
	\begin{fleqn}  
		\begin{equation}  \small
		\begin{aligned}[b]
		{J_i}(k)&={{\lVert{x_i(K)}\lVert}_{P_i}^2}+  ( {\lVert{({x_i(k)}-{{x_i}^d}(k))\lVert}_{Q_i}^2} + {\lVert{{z_i(k)}\lVert}_{S_i}^2}+{\lVert{(U(k))\lVert}_{R_i}^2}) 
		\label{objectivei}
		\end{aligned}
		\end{equation}
	\end{fleqn}   \indent
	\normalsize
	Note that the control variables $T_{ac}i$, $M_{ac}i$, $wf_{ac}i$, and the energy consumption are considered in controller $i$ of room $i$. Furthermore, the subsystems' interactions are included in the objective functions using the Goal coordination algorithm [5], such that the subsystems' dynamic constraints with weighting coefficients are imposed in the cost function. Moreover, each state or input constraint can be regarded separately in the algorithm using the diagonal weighting matrices.
	
	\begin{fleqn}  
		\begin{equation} \small
		\begin{aligned}[b]
		{M_i}(P) &= min \{ {{\lVert{x_i(K)}\lVert}_{P_i}^2}+ \sum\limits_{k=0}^{K-1} ( {\lVert{({x_i(k)}-{{x_i}^d}(k))\lVert}_{Q_i}^2} + {\lVert{{z_i(k)}\lVert}_{S_i}^2}\\
		&+ {\lVert{({T_{ac}}(k))\lVert}^2}_{R_i}  + {{P_i}^T}(k)[A_ix_i(k)+B_iu_i(k)+C_iz_i(k)\\
		&-x_i(k+1)]+  {\lambda _i}^T(z_i-\sum\limits_{j=1}^{m} L_{ij}x_j)             )  \}             
		\label{tamuraMp}
		\end{aligned}
		\end{equation}
	\end{fleqn}
	\\
	\normalsize  
	where $L_{ij}$s are the coefficients used for connecting the states of neighboring subsystems. $\lambda _i$ and  ${P_i}^T$ are the interactions and system model constraints respectively. Using the Hamiltonian function $H_i$ defined in (\ref{tamuraH}), (\ref{tamuraMp}) can be converted to (\ref{tamuraMp_revised}). \\
	
	\begin{fleqn}  \small
		\begin{equation} 
		\begin{aligned}[b]
		{H_i}(x_i, u_i, z_i, k) &= {\lVert{({x_i(k)}-{{x_i}^d}(k))\lVert}_{Q_i}^2} + {\lVert{{z_i(k)}\lVert}_{S_i}^2}+ {\lVert{({T_{ac}}(k))\lVert}_{R_i}^2} \\ 
		&+ {{P_i}^T}(k)[A_ix_i(k)+B_iu_i(k)+C_iz_i(k)-x_i(k+1)]\\
		&+  {\lambda _i}^T(z_i-\sum\limits_{j=1}^{m} L_{ij}x_j)                  
		\label{tamuraH}
		\end{aligned}
		\end{equation}
	\end{fleqn}
	\begin{fleqn}   \small
		\begin{equation} 
		\begin{aligned}[b]
		{M_i}(P) &= min \{ {{\lVert{x_i(K)}\lVert}_{P_i}^2} - {{{P_i}(K-1)}^T} {x_i(K)} \\
		&+ \sum\limits_{k=0}^{K-1} (H_i(k)-P_i(k-1)x_i(k)) )  \}             
		\label{tamuraMp_revised}
		\end{aligned}
		\end{equation}
	\end{fleqn}
	
	\normalsize
	In each instant, the following proposed three-level algorithm is being applied iteratively up to the prediction horizon until the optimum input is attained.   
	\begin{itemize}
		\item k=0; minimize $H_i(x_i(0),u_i(0),z_i(0))$ with partial derivatives with respect to $u_i(0)$ and $z_i(0)$.
		\item k=1,2,...,K-1; minimize $H_i(x_i(k),u_i(k),z_i(k),k)-{{P_i}(k-1)}^Tx_i(k))$ with respect to $x_i(k)$, $u_i(k)$ and $z_i(k)$.
		\item k=K; minimize ${\lVert{({x_i(K)})\lVert}_{P_i}^2}-{{P_i}^T}(K-1)x_i(K)$ with respect to $x_i(K)$. 
	\end{itemize} 
	
\begin{figure}[thpb]
		\begin{center}
			\framebox{\parbox{3in}{
					\centering
					\includegraphics[height=5.5cm, width=7.2cm]{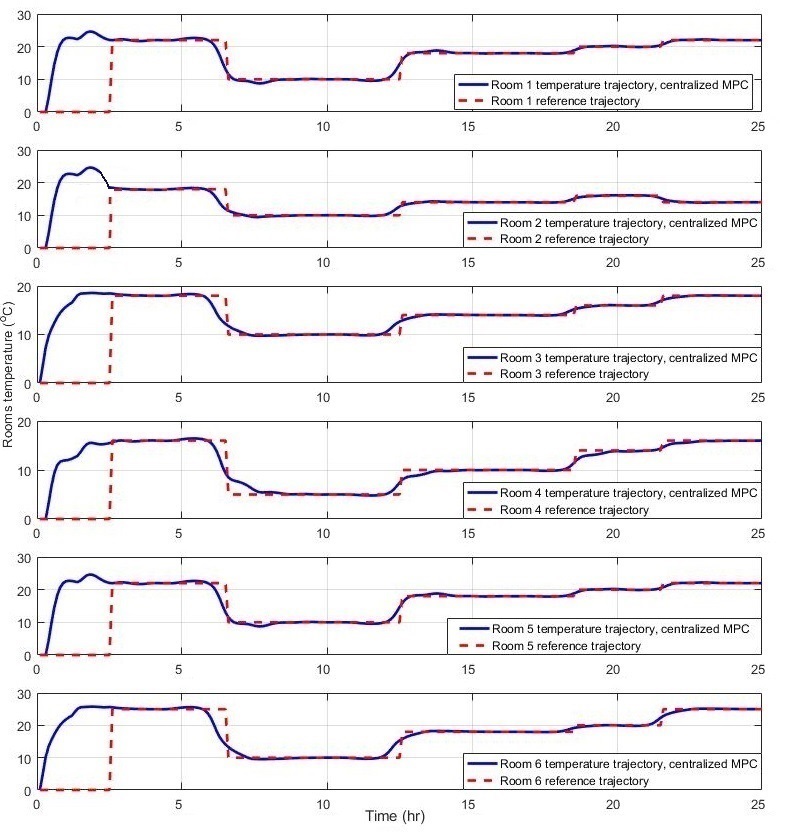}
			}}
			\caption{Six rooms' temperature using centralized MPC}
			\label{fig:T1distributed-centralized}
		\end{center}
		
		\begin{center}
			\framebox{\parbox{3in}{
					\centering
					\includegraphics[height=5.5cm, width=7.2cm]{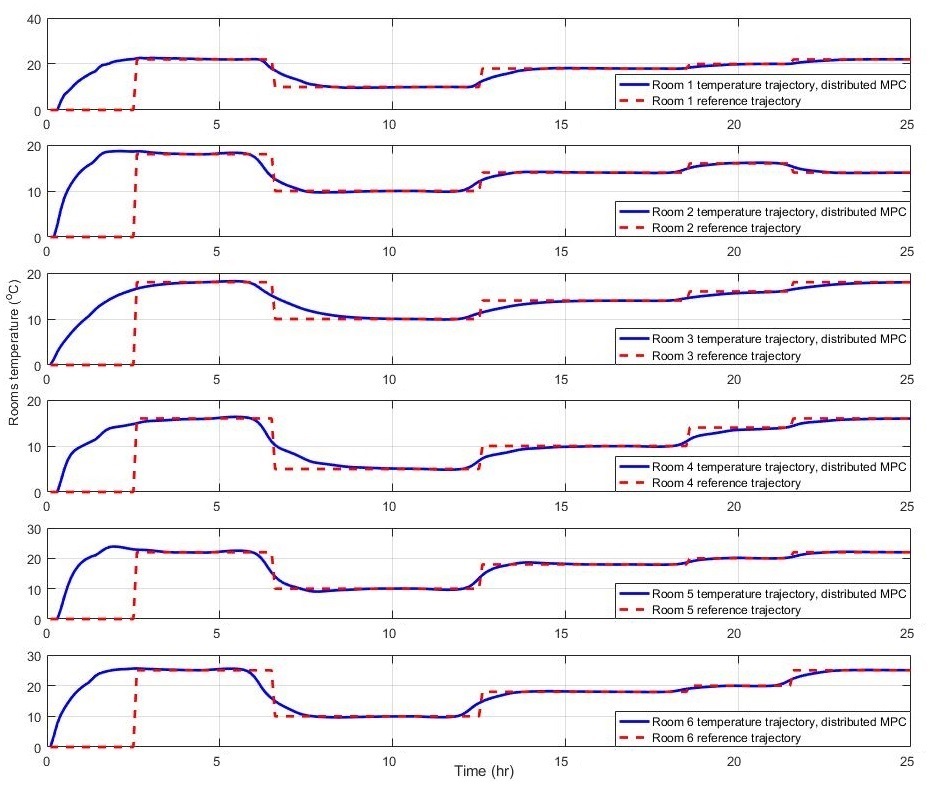}
			}}
			\caption{Six rooms' temperature using distributed MPC}
			\label{fig:T2distributed-centralized}
		\end{center}
		
		\begin{center}
			\framebox{\parbox{3in}{
					\centering
					\includegraphics[height=3.5cm, width=7.2cm]{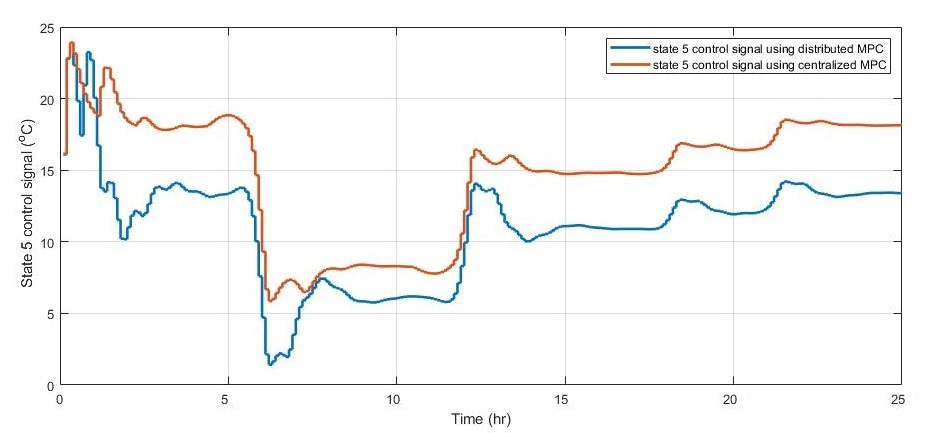}
			}}
			\caption{Control signal 5 using centralized and distributed MPC}
			\label{fig:Tac_distributed-centralized_2}
		\end{center}
	\end{figure}

	\normalsize
	Fig.s \ref{fig:T1distributed-centralized} and \ref{fig:T2distributed-centralized} show the six rooms' temperatures using the centralized and distributed MPC respectively. Fig. \ref{fig:Tac_distributed-centralized_2} illustrates the control trajectory results from the centralized and distributed MPC respectively. Table \ref{tab:numerical charsteristics} compares the numerical values of the two rooms' temperatures and their control trajectory characteristics using  centralized and distributed MPC cases.  
	\noindent
	\noindent
	\begin{table}  
		\begin{center}  
			\caption{Numerical characteristics of the states and control plots using centralized and distributed MPC}
			\small
			\begin{tabular}{| m{0.7cm} | m{0.7cm} | m{0.7cm} | m{0.7cm} | m{0.7cm} | m{0.7cm} | m{1.2cm} | m{0.7cm} |}
				\hline
				\tiny
				& {$T_1$ overshoot} \tiny & {$T_1$ peak value} \tiny & {$T_2$ overshoot} \tiny & {$T_2$ peak value} \tiny & {control overshoot} \tiny & {control area} \tiny & {run time} \tiny \\  \hline 
				CMPC \tiny & 89.95 \tiny & 22.39 \tiny & 80.96 \tiny & 25.03 \tiny & 78.66 \tiny &  7.4138e+3 \tiny & 3120 sec \tiny \\ \hline  
				DMPC \tiny & 11.65 \tiny & 20.33 \tiny & 15.70 \tiny & 17.28 \tiny & 31.90 \tiny &  5.5291e+3 \tiny & 52 sec \tiny \\ \hline
			\end{tabular}
			\label{tab:numerical charsteristics}
			\normalsize
		\end{center}
	\end{table}  
	\indent
	\normalsize
	From Fig.s \ref{fig:T1distributed-centralized} and \ref{fig:T2distributed-centralized}, the distributed MPC functions better compared with the centralized MPC strategy in terms of reference tracking performance. From Table \ref{tab:numerical charsteristics}, the overshoots and peak values of room 1 and 2 temperatures using distributed MPC are significantly smaller than the same values in centralized case. \\ \indent
	From Fig. \ref{fig:Tac_distributed-centralized_2}, the distributed MPC control signal shows lower overshoot and stabilizes sooner than the centralized MPC control trajectory. Comparing the areas under the control signals (Table \ref{tab:numerical charsteristics} 7th column), the energy consumption using distributed MPC is 25.42\% lower than that of the centralized one. Besides, the optimization time using the DMPC and centralized MPC in a Corei7, 3.2GHz computer are shown in the last column of Table \ref{tab:numerical charsteristics}. Hence, DMPC controller performs 60 times faster than the centralized MPC. As the plant gets larger, the computation time using the centralized framework gets relatively high which makes the real-time control of the system impossible. Another important innovation of the proposed DMPC algorithm is that it considers the disturbances predictions and it owns stability and feasibility properties. The implemented controlled system is being applied in an IOT project. In fact, using the proposed scheme for the IOT building sector, not all the agents need to be connected to each other, therefore the communication effort is significantly lower compared to the centralized scheme.
	
	\section{CONCLUSIONS}
	A distributed MPC and a centralized MPC strategy have been developed for a benchmark temperature control problem of a six-zone building system. The aim was to regulate the building's six zones' thermal condition to the desired set-points with minimum energy consumption. The heat exchange between the rooms, and between the outer and inner spaces are all considered in the system model. The control variables are the heat flow amount and the heater temperature in the zones. The proposed distributed predictive controller is able to predict the model inputs, states, thermal exchanges, and disturbances to compensate the system outputs rapidly.\\ \indent
	From the simulation results, the distributed MPC approach showed better performance in signal tracking, energy consumption (25.42\% less), and computation time (60 times lower) compared with the centralized MPC. The proposed distributed MPC improved control performance by utilizing the disturbances predictions as part of the system model. Besides, the stability of the closed-loop system using the proposed distributed algorithm is guaranteed through the Lyapunov theory. Moreover, the feasibility of the solution is guaranteed if the initial solution is feasible and the controlled closed-loop system is asymptotically stable at the system's equilibrium point. The distributed MPC method proposed here can be generalized to a larger plant of this kind for future works. 
	

\end{document}